# Crystals of Na$^+$ ions at the surface of a silica hydrosol

Aleksey M. Tikhonov*

*University of Chicago, Consortium for Advanced Radiation Sources, and Brookhaven National Laboratory, National Synchrotron Light Source, Beamline X19C, Upton, NY, 11973, USA*

**January 5, 2007**

I used x-ray grazing incidence diffraction to measure the spatial correlations between sodium ions adsorbed with Bjerrum's density at the surface of a monodispersed 22-nm-particle colloidal silica solution stabilized by NaOH with a total bulk concentration ~ 0.05 mol/L. My findings show that the surface compact layer is in a two-dimensional crystalline state (symmetry $p2$), with four ions forming the unit cell and a ~30 Å translational correlation length between sodium ions.

The symmetry of the free surface of a liquid can be anisotropic, regardless of the bulk's isotropic symmetry.[1] With a density of adsorbed ions at the liquid's surface as high as $\sim R_B^{-2}$ (Bjerrum's density), the average energy of their interactions is comparable to their thermal energy $\sim k_B T$, where $k_B$ is Boltzmann's constant and $T$ is temperature. At room temperature, the Bjerrum radius for monovalent ions in aqueous media, $R_B$, is $\sim 7$ Å. According to Vorotyntsev and Ivanov,[2] at Bjerrum's density ($>10^{18}$ m$^{-2}$), the ions can transit into a two-dimensional condensed phase so that the surface symmetry may become anisotropic. In this letter, I discuss my study, using X-ray grazing incidence diffraction, of the spatial correlations between alkali ions adsorbed at the surface of a monodisperse colloidal silica solution, wherein the average distance between them, $\xi \sim 5$ Å, is less than $R_B$. According to my data, the surface compact layer is in a two-dimensional crystalline state with a translational correlation length of $\sim 30$ Å between sodium ions. As far as I am aware, this is the first report on the crystallization of two-dimensional ions adsorbed with Bjerrum's density at the free surface of a liquid.

My recent x-ray scattering studies of the surface of the nanocolloidal silica solutions stabilized by NaOH (pH $\sim$ 9-10) revealed a spontaneously formed non-trivial surface-normal structure that reflects a colossal difference in the potentials of "image forces" for cationic alkali ions and anionic nanoparticles (Fig 1a).[3,4] At room temperature, the structure consists of a 2-nm-thick compact layer of Na$^+$ with a surface concentration $\Gamma^+ \sim 10^{19}$ m$^{-2}$, and an underlying loose monolayer of nanocolloidal particles as part of a thick diffuse layer, in between which is a layer with a low concentration of electrolytes (water layer).[5] According to my previous x-ray reflectivity study, the compact layer can be described by a two-layered model, i.e., a $\sim 6$-Å-thick low-density layer (layer 1) of directly adsorbed hydrated alkali ions with a surface concentration $\Gamma_1^+ \sim 3\times10^{18}$ m$^{-2}$, and a $\sim 13$-Å-thick high-density layer (layer 2) with a surface concentration of sodium ions $\Gamma_2^+ \sim 8\times10^{18}$ m$^{-2}$.[4] The main difference between them is their alkali ion/water ratio: the number of H$_2$O molecules per ion in the low-density layer 1 with a total surface density of electrons in it, $\Gamma_1 \sim 10^{20}$ m$^{-2}$, is as much as $(\Gamma_1/10 - \Gamma_1^+)/\Gamma_1^+ \sim 2$ (both Na$^+$ and H$_2$O contain 10 electrons); however, the number of H$_2$O molecules per one Na$^+$ in layer 2 is much larger $\left(\Gamma_2/10 - \Gamma_2^+\right)/\Gamma_2^+ \sim 10$, where $\Gamma_2 \sim 7\times 10^{20}\,\mathrm{m}^{-2}$ is the total surface density of electrons in this layer. Therefore, the dielectric permittivity, $\varepsilon$, in layer 1 must be approximately five times less than that in layer 2, so that the interaction between ions in layer 1 is stronger ($R_B \sim 1/\varepsilon$) than in layer 2, since $\xi \approx \left(l_1/\Gamma_1^+\right)^{1/3} \approx \left(l_2/\Gamma_2^+\right)^{1/3} \sim 5-6\,\text{Å}$ is roughly the same for both layers.

I explored the surface of a silica sol with 22-nm-particles to help interpret the diffraction data. Advantageously, the penetration length, $\Lambda$, (~ 10 nm) of the x-rays in the grazing-incident diffraction experiment at the sol's surface was noticeably smaller than the width of the surface transitional region: the plane of the closest approach of the nanoparticles to the surface lies ~15 nm beneath it, such that a thick layer of water separates the compact layer well from the silica particles.

X-ray grazing incidence diffraction is the standard technique for looking at the in-plane structure at solid and liquid surfaces with atomic spatial resolution.[6-8] I studied the in-plane structure of eight liquid samples using a synchrotron x-ray liquid surface spectrometer at beamline X19C, National Synchrotron Light Source, Brookhaven National Laboratory[9], employing a monochromatic focused x-ray beam ($\lambda = 0.825 \pm 0.002$ Å) to explore the hydrosol's planar surface. Samples of colloidal silica (Ludox TM 40, supplied by Grace Davison)[10] in a ~ 100 ml capacity glass-dish with a circular interfacial area (100 mm diameter) were equilibrated at $T = 298$ K inside a vapor-tight single-stage thermostat and mounted above the level of water in the bath (~200 mm diameter), which served as a humidifier. The surface tension of sol's surface was as large as 74 mN/m, as measured by the Wilhelmy plate method.

The kinematics of scattering at the liquid's surface can be advantageously described within a right-handed rectangular coordinate system wherein the origin, $O$, is in the center of the x-ray footprint; here, the $xy$ plane coincides with the air/sol interface, the axis $x$ is perpendicular to the beam's direction, and the axis $z$ is directed normal to the interface opposite to the gravitational force. In the insert in Fig. 1b, $\alpha$ is the incident angle in the $xz$ plane, $\beta$ is the angle in the vertical plane between the scattering direction and the interface, and $\phi$ is the angle in the $xy$ plane between the incident beam's direction and the direction of scattering. At incident angles below critical, $\alpha_c \approx 0.09$ deg, a monochromatic 15 keV x-ray beam is totally reflected from the sol's surface.[4] At $\alpha < 0.8\alpha_c$, the penetration depth, $\Lambda$, of the x-rays is very shallow and scattering occurs in the top $\sim \lambda/(2\pi\alpha_c) \sim 100$ Å thick layer, where the intensity is usefully expressed as a function of the in-plane component of the wave-vector transfer $q_{xy} = (q_x^2 + q_y^2)^{1/2}$.[8] At $\alpha, \beta << 1$, $q_{xy} \approx (4\pi/\lambda)\sin(\phi/2)$.

Fig. 2 depicts the intensity of scattering from the surface of the hydrosol that contains three diffraction peaks. It was recorded at a grazing angle $\alpha \approx 0.07$ deg with a vertical position-sensitive detector (Ordela) by summing 10 channels covering the range of $\Delta\beta \sim 0.1$ deg ($\Delta q_z \sim 2\times 10^{-2}$ Å$^{-1}$) at $q_z \approx 0.08$ Å$^{-1}$. In this experiment, the incident beam's vertical size, set by input slits, was as large as $\sim 40\,\mu$m, and the horizontal resolution of the detector, set by Soller slits, was as high as $\Delta_{xy} \approx 0.02$ Å$^{-1}$ ($\Delta\phi \approx 0.18$ deg). Fig 3 illustrates the distribution of intensity of grazing-incidence scattering in the $q_z$ vs. $q_{xy}$ plane where the diffraction peaks were recorded by summing $\sim 400$ channels divided into $\sim 30$ groups (the number of horizontal lines in the image).

The diffraction pattern in Fig. 3 can be qualitatively explained based on the surface-normal model shown in Fig. 1a. The structure factor of such an "interfacial sandwich" is a set of ''Bragg rods'' normal to the surface plane at $q_z^2 << q_{xy}^2$.[6-8] Fortunately, bulk water and the water's surface scatter x-rays very similarly. For example, the grazing-incidence scattering from the surface of water at $q_z = 0$ and the intensity of scattering from bulk water have a broad diffraction peak at $\sim 2$ Å$^{-1}$ associated mainly with the O-O correlations in water at distances $\sim 3$ Å.[4, 11] The diffraction peaks in Fig. 2 correspond to the in-plane correlations at noticeably larger distances ($\sim 4$ Å) and cannot be related to scattering in the water layer. On the other hand, the average distance between ions in layers 1 and 2 is $\xi \sim 5 - 6$ Å, so that the narrow diffraction peaks could reflect two-dimensional crystallization of the hydrated Na$^+$ in them.

The diffraction pattern can be described by an oblique two-dimensional Bravais lattice (symmetry $p2$) defined by two planar vectors $\boldsymbol{a}_1$ and $\boldsymbol{a}_2$ of the lattice translations and an angle, $\gamma$, between them, with four Na$^+$ forming a unit cell (Fig. 4).[12] Since the distance between first two peaks at $q_{xy}^{10} \approx 1.5$ Å$^{-1}$ and $q_{xy}^{01} \approx 1.7$ Å$^{-1}$ is relatively small, the lattice deviates slightly from the square net ($a_1 = a_2 = a$, $\gamma = \pi/2$). If the splitting of the (10)-reflection, $\kappa = a\left(q_{xy}^{01} - q_{xy}^{10}\right)/2\pi \approx 0.1$, for the square lattice with period $a = 4\pi/\left(q_{xy}^{10} + q_{xy}^{01}\right) \approx 3.9$ Å, then, in accordance with Bragg-Wulff law, the first two rods immediately denote the distances between ionic rows with the lowest indexes in the lattice $d_{10} \approx a(1 + (1/2)\kappa) \approx 4.1$ Å and $d_{01} \approx a(1 - (1/2)\kappa) \approx 3.7$ Å. The position of the third peak at $q_{xy}^{11} \approx 2.07$ Å$^{-1}$ ($d_{11} = 2\pi/q_{xy}^{11} \approx 3.0$ Å) differs from the position of the (11)-reflection for the square lattice ($= 2\sqrt{2}\pi/a \approx 2.3$ Å$^{-1}$). It can be related to a small deviation of $\gamma$ from $\pi/2$ that is established from the equation for the area of the triangle $ABC$, $(1/2)ACd_{11} = (1/2)a_2 d_{10}$ (see Fig. 4). If $\delta = \pi/2 - \gamma \ll 1$, then $AC \approx \sqrt{2}a(1 - \delta/2)$. Since $a_2 d_{01} \approx a^2$, then $\delta \approx 2 - \sqrt{2}(a/d_{11}) \approx 0.2$ so that $a_1 \approx d_{10} \approx 4.1$ Å, $a_2 \approx d_{01} \approx 3.7$ Å and $\gamma \approx 80$ deg. However, the $(1\bar{1})$-reflection at $q_{xy}^{1\bar{1}} \sim 2.5$ Å$^{-1}$ ($d_{1\bar{1}} \approx a(1 - \delta/2)/\sqrt{2} \approx 2.5$ Å) was not observed, possibly due to lattice fluctuations that usually suppress the intensities of diffraction peaks at large $qs$.[12, 13]

In Fig. 2 all diffraction peaks have roughly the same width, $\Delta Q_{xy} \sim 0.05$ Å$^{-1}$ along $q_{xy}$, which is noticeably larger than the in-plane resolution of the experiments, i.e., $\Delta_{xy}$. Their width signifies that the translational correlation length, $L$, between sodium ions in the crystal is as large as $L \sim a^2/\Delta a \sim \pi(a/\Delta Q)^{1/2} \approx 30$ Å, where $\Delta Q = \left(\Delta Q_{xy}^2 - \Delta_{xy}^2\right)^{1/2}$, and $\Delta a \sim a^{3/2}\Delta Q^{1/2}/\pi \sim 0.5$ Å is the average deviation of an ion from its site in the lattice.[13] I suggest that long-range-order is destroyed by spatial fluctuations in the interfacial electric field due to the finite size of the silica nanoparticles so that the diffraction pattern seen in Figs 2 and 3 is averaged over all crystallites inside the beam's footprint (so-called ''powder averaging''). I also note that $\Delta a$ explains the large range of $\Delta q_{xy}$ where, for example, the (10)-reflection is observed: Laue's diffraction conditions for $\mathbf{q}$ are fulfilled within the range, $\Delta q_{xy} \approx 2\pi \Delta a/a^2 \sim 0.2$ Å$^{-1}$.

The area of the unit cell in an oblique model lattice is as large as $d_{10}d_{01} \approx 15$ Å$^2$, i.e., it can accommodate one water molecule. Reasonably, I suggest that the projection of the oxygen atom of the water molecule in the unit cell is centered at the intersection of the diagonals *AC* and *BD* in Fig. 4: water molecules form the same oblique lattice as does Na$^+$. Although the symmetry of the lattice *p*2 reflects the anisotropy of the water molecules (point group *mm*2), their exact orientation in the crystal was not established; it requires quantitative analysis of the diffraction data.

The surface density of ions in the crystal, $\Gamma_0^+ \approx 6 \times 10^{18}$ m$^{-2}$, is twice as large as $\Gamma_1^+$ in layer 1 (the layer with the strongest ion-ion in-plane interaction) as established from a reflectivity study of cesium-enriched sols, and is comparable to the density of ions in layer 2, $\Gamma_2^+$, which points to the crystallization of layer 2. Alternatively, the surface may only be partially covered by the crystallites in layer 1 (~ 6 nm in diameter), so that the total surface charge density associated with the Na$^+$ crystal would be lower then $\Gamma_0^+$. Unfortunately, all layers scatter X-rays coherently, so that separating the layers' contributions to the scattering intensity based on the present data is a complicated problem. In further experiments I plan to investigate the dependence of the in-plane structure of the compact layer on the size of the silica particles in the hydrosol, which could help in revealing precisely the positioning of the crystals in the compact layer from the alterations in the mosaicity of the surface.

Finally, ions adsorbed at the silica hydrosol's surface can be considered as a heavy, very dense analog of the two-dimensional system of electrons suspended above the surfaces of certain cryogenic dielectrics (liquid $^3$He and $^4$He, and liquid and solid hydrogen) by the image force and an external electrostatic field.[14,15] The energy, $W$, of an ion in the 2D lattice, defined by Coulomb interaction with the ions within the correlation length (~$10^2$) in the lattice, is roughly as high as $W \sim 30 e^2 / (4\pi\varepsilon_0 \varepsilon a)$ ~ $200 k_B T - 400 k_B T$ (where $e$ is elementary charge, $\varepsilon_0$ is the dielectric permittivity of the vacuum, and $\varepsilon \sim 10$), which is comparable with the hydration energy of Na$^+$ in water (~$160 k_B T$). Others previously observed a solid phase of "classical" two-dimensional electrons (a Wigner crystal) at the surface of liquid helium when the ratio of potential- to kinetic-energy per electron becomes more than ~ 140, i.e., comparable to $W/k_B T$.[16-17] In those experiments the temperature was much lower (~ 0.5 K) than in mine because the experimentally accessible range of surface densities of electron gases, for example, at the surface of liquid helium, is $\sim 10^4$ times smaller than the density of sodium ions at the sol's surface.

## ACKNOWLEDGEMENTS

Use of the National Synchrotron Light Source, Brookhaven National Laboratory, was supported by the U.S. Department of Energy, Office of Science, Office of Basic Energy Sciences, under Contract No. DE-AC02-98CH10886. X19C is partially supported through funding from the ChemMatCARS National Synchrotron Resource, the University of Chicago, the University of Illinois at Chicago and Stony Brook University. I thank Vladimir I. Marchenko and Alexander G. Abanov for valuable discussions. The author also thanks Avril Woodhead for her comments on the manuscript.

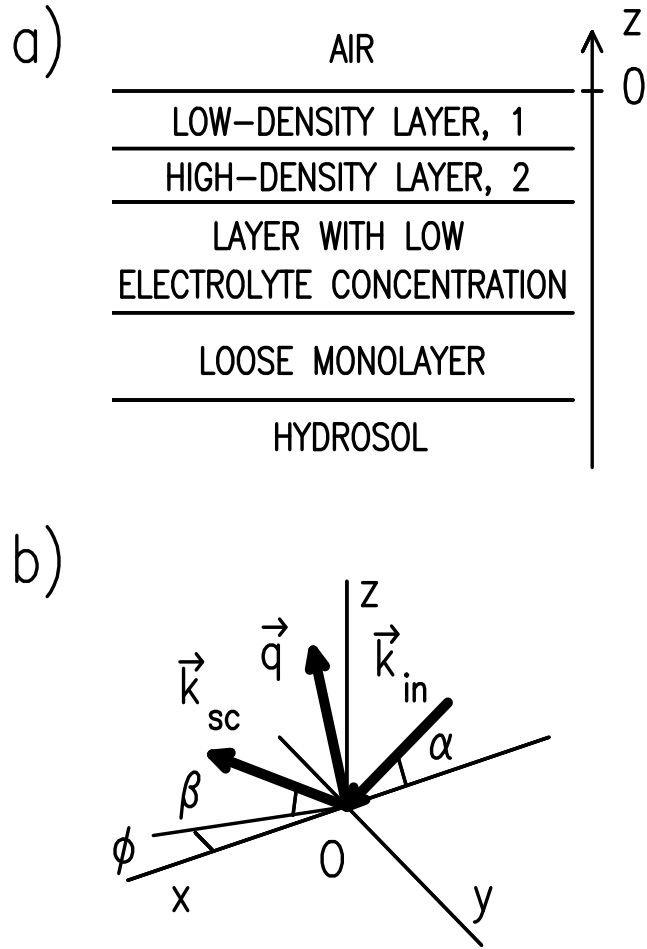

**Figure 1**. a) The surface-normal structure of the silica hydrosol's surface. It consists of a 2-nm thick compact layer of $Na^+$ with the surface concentration, $\Gamma^+$, as large as $\sim 10^{19}\,m^{-2}$, a loose monolayer of nanocolloidal particles ($\sim$ 20-nm-thick) as part of a thick diffuse layer, in between which is a $\sim$15-nm-thick layer with low electrolyte concentration. The compact layer can be described by a two-layer model, i.e., a $\sim$ 6-Å-thick low-density layer (layer 1) of directly adsorbed hydrated alkali ions, and a $\sim$ 13-Å-thick high-density layer (layer 2).[4] b) Sketch of the kinematics of the scattering at the silica sol's surface. The *x-y* plane coincides with the interface, the axis *y* is perpendicular to the beam's direction, and the axis *z* is directed normal to the surface opposite to the gravitational force. $\mathbf{k}_{in}$ and $\mathbf{k}_{sc}$ are, respectively, wave-vectors of the incident beam, and beam scattered towards the point of observation, and $\mathbf{q}$ is the wave-vector transfer, $\mathbf{q} = \mathbf{k}_{in} - \mathbf{k}_{sc}$.

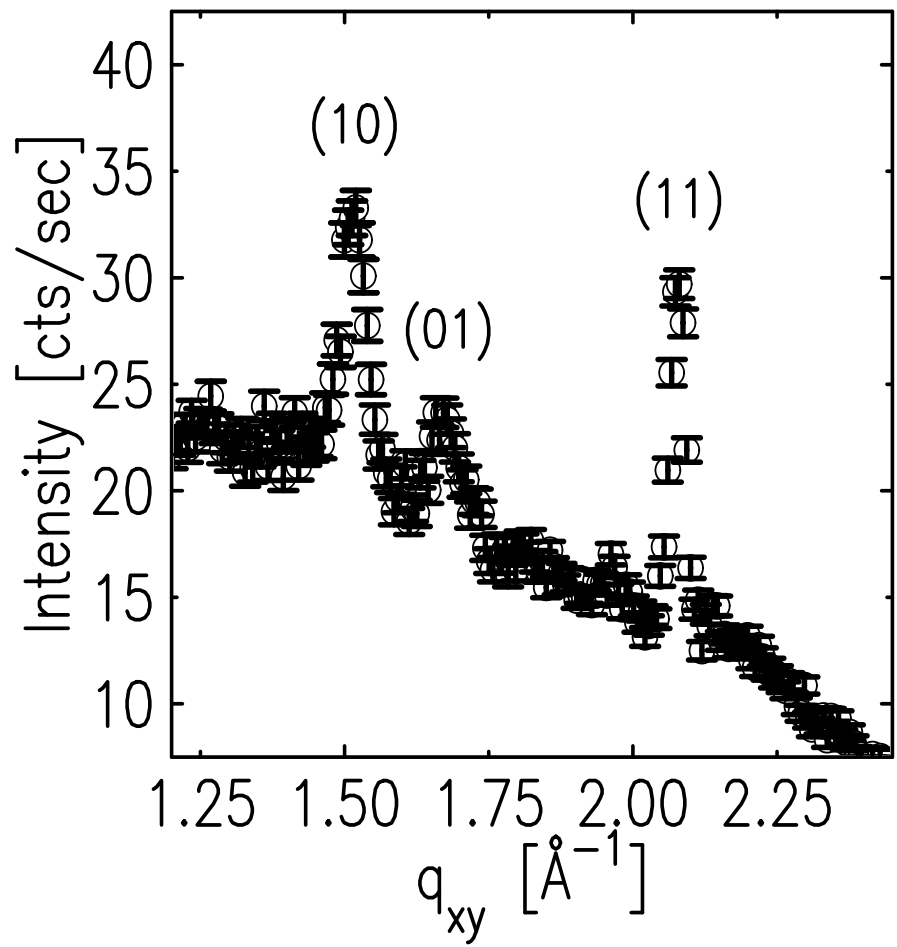

**Figure 2.** The intensity of grazing-incidence diffraction at ambient conditions from the surface of the aqueous solution of a suspension of 22-nm silica particles. It was recorded at a glancing angle $\alpha \approx 0.07$ deg with a vertical position-sensitive detector (Ordela) by summing ten channels over a range of $\Delta\beta \sim 0.1$ deg ($\Delta q_z \approx 0.02$ Å$^{-1}$) at $q_z \approx 0.08$ Å$^{-1}$. The detector's horizontal resolution, set by the Soller slits, was as much as $\Delta q_{xy} \approx 0.02$ Å$^{-1}$ ($\Delta\phi \approx 0.18$ deg).

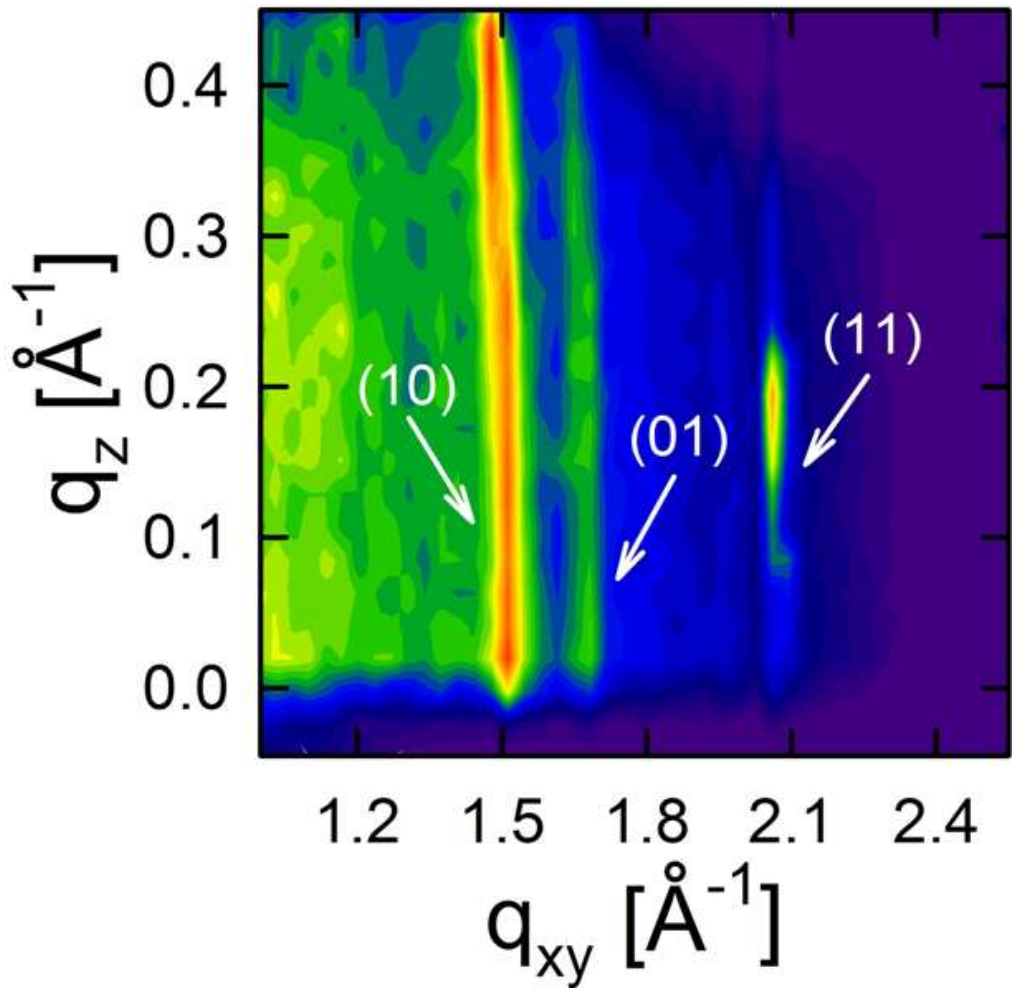

**Figure 3.** The intensity of grazing-incidence scattering in the $q_z$ vs. $q_{xy}$ plane. It was recorded at a grazing angle of $\alpha \approx 0.07$ deg summing ~ 400 channels of the position-sensitive detector divided into ~ 30 groups (number of horizontal lines in the image).

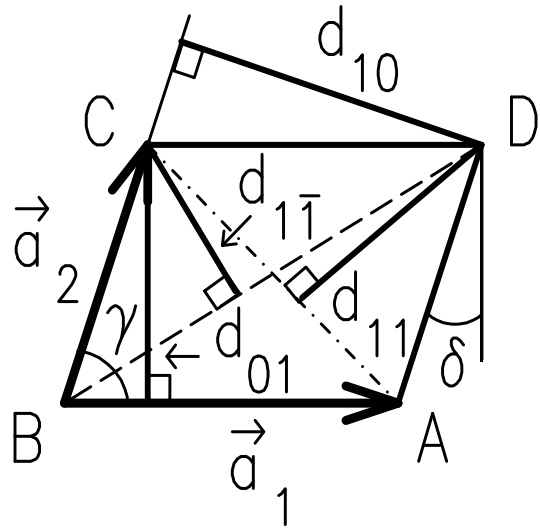

**Figure 4.** The oblique unit cell of the two-dimensional Bravais lattice of Na$^+$ (symmetry $p2$) at the hydrosol's surface: $a_1 \approx 3.7$ Å, $a_2 \approx 4.1$ Å and $\gamma \approx 80$ deg. Sodium ions are centered at the corners of the parallelogram $ABCD$, the dashed line ($AC$) is the $(1\bar{1})$-row and dash-dotted ($BD$) line is $(11)$-row. A small deviation of $\gamma$ from $\pi/2$, $\delta$, can be established, for example, from the equation for the area of the triangle $ABC$, $(1/2)ACd_{11} = (1/2)a_2 d_{10}$.